\documentclass[reqno]{amsart}
\usepackage{color}
\usepackage{amsmath,amsthm}
\usepackage{amsfonts}

\usepackage[all]{xy}
\SelectTips{cm}{10}

\usepackage{graphicx}

\usepackage{tikz}
\usetikzlibrary{backgrounds,circuits,circuits.ee.IEC,shapes,fit,matrix}
\usetikzlibrary{arrows,positioning}
\tikzstyle{main node} =[circle,fill=white!20,draw,font=\sffamily\Large\bfseries]
\tikzstyle{terminal}=[circle,fill=white!20,draw,font=\sffamily\Large\bfseries,color=purple,fill=none]

\definecolor{myurlcolor}{rgb}{0.5,0,0}
\definecolor{mycitecolor}{rgb}{0,0,0.8}
\definecolor{myrefcolor}{rgb}{0,0,0.8}
\usepackage[pagebackref,bookmarks=false]{hyperref}
\hypersetup{colorlinks,
linkcolor=myrefcolor,
citecolor=mycitecolor,
urlcolor=myurlcolor}

\newcommand{\N}{\mathbb{N}}

\newcommand{\R}{\mathbb{R}}

\newcommand{\maps}{\colon}
\newcommand{\beq}{\begin{equation}}
\newcommand{\eeq}{\end{equation}}

\theoremstyle{plain}

\theoremstyle{definition}

\theoremstyle{remark}

\begin{document}



\title{Relative entropy in biological systems}

\author{John C.~Baez}
\address{Department of Mathematics\\ 
University of California\\ 
Riverside CA 92521\\
USA \\
and Centre for Quantum Technologies\\ 
National University of Singapore\\ 
Singapore 117543}

\author{Blake S.~Pollard}
\address{Department of Physics and Astronomy\\ 
University of California\\ 
Riverside CA 92521\\
USA}

\date{February 10, 2016}

\begin{abstract}
In this paper we review various information-theoretic characterizations of the approach to equilibrium in biological systems.  The replicator equation, evolutionary game theory, Markov processes and chemical reaction networks all describe the dynamics of a population or probability distribution.  Under suitable assumptions, the distribution will approach an equilibrium with the passage of time.  Relative entropy---that is, the Kullback--Leibler divergence, or various generalizations of this---provides a quantitative measure of how far from equilibrium the system is.  We explain various theorems that give conditions under which relative entropy is nonincreasing.   In biochemical applications these results can be seen as versions of the Second Law of Thermodynamics, stating that free energy can never increase with the passage of time.   In ecological applications, they make precise the notion that a population gains information from its environment as it approaches equilibrium.
\end{abstract}

\maketitle

\section{Introduction}
\label{introduction}

Life relies on nonequilibrium thermodynamics, since in thermal equilibrium there are no flows of free energy.  Biological systems are also open systems, in the sense that both matter and energy flow in and out of them.  Nonetheless, it is important in biology that systems can sometimes be treated as approximately closed, and sometimes approach equilibrium before being disrupted in one way or another.   This can occur on a wide range of scales, from large ecosystems to within a single cell or organelle.  Examples include:
\begin{itemize}
\item 
a population approaching an evolutionarily stable state;
\item 
random processes such as mutation, genetic drift, the diffusion of organisms in an environment or the diffusion of molecules in a liquid;
\item 
a chemical reaction approaching equilibrium.
\end{itemize}

A common feature of these processes is that as they occur, quantities mathematically akin to entropy tend to increase.  Closely related quantities such as free energy tend to decrease.  In this review we explain some mathematical results that explain why this occurs.  

Most of these results involve a quantity that is variously known as `relative
information', `relative entropy', `information gain' or the `Kullback--Leibler divergence'.  We shall use the first term.   Given two probability distributions $p$ and $q$ on a finite set $X$, their {\bf \boldmath{relative information}}, or more precisely the {\bf \boldmath{information of $p$ relative to $q$}}, is
\beq        \displaystyle{     I(p,q) = \sum_{i \in X} p_i \ln\left(\frac{p_i}{q_i}\right) . } \label{relative_information} \eeq
We use the word `information' instead of `entropy' because one expects entropy to increase with time, and the theorems we present will say that $I(p,q)$ decreases with time under various conditions.   The reason is that the Shannon entropy
\beq    \displaystyle{   S(p) = -\sum_{i \in X} p_i \ln p_i } \label{shannon_entropy}  \eeq
contains a minus sign that is missing from the definition of relative information.  

Intuitively, $I(p,q)$ is the amount of information gained when we start with a hypothesis given by some probability distribution $q$ and then change our hypothesis, perhaps on the basis of some evidence, to some other distribution $p$.   For example, if we start with the hypothesis that a coin is fair and then are told that it landed heads up, the relative information is $\ln 2$, so we have gained 1 bit of information.  If however we started with the hypothesis that the coin always lands heads up, we would have gained no information.

Mathematically, relative information is a {\bf \boldmath{divergence}}: it obeys
\[       I(p, q) \ge 0 \]
and 
\[       I(p, q) = 0 \iff p = q \]
but not necessarily the other axioms for a distance function, symmetry and the triangle inequality, which indeed fail for relative information.  There are many other divergences besides relative information \cite{Crooks,GorbanGorbanJudge}.  However, relative information can be singled out by a number of characterizations \cite{Hobson}, including one based on ideas from Bayesian inference \cite{BaezFritz}.  The relative information is also close to the expected number of extra bits required to code messages distributed according to the probability measure $p$ using a code optimized for messages distributed according to $q$ \cite[Theorem 5.4.3]{CoverThomas}.

In this review we describe various ways in which a population or probability distribution  evolves continuously according to some differential equation.  For all these differential equations, we describe conditions under which relative information decreases.  Briefly, the results are as follows.  We hasten to reassure the reader that our paper explains all the terminology involved, and the proofs of the claims are given in full:

\begin{itemize}
\item
In Section \ref{sec.replicator} we consider a very general form of the Lotka--Volterra equations, which are a commonly used model of population dynamics.   Starting from the population $P_i$ of each type of replicating entity, we can define a probability distribution
\[            p_i  = \frac{P_i}{\displaystyle{\sum_{i \in X} P_i }} \]
which evolves according to a nonlinear equation called the replicator equation.  We describe a necessary and sufficient condition under which $I(q,p(t))$ is nonincreasing when $p(t)$ evolves according to the replicator equation while $q$ is held fixed.  
\item 
In Section \ref{game} we consider a special case of the replicator equation that is widely
studied in evolutionary game theory.   In this case we can think of probability 
distributions as mixed strategies in a two-player game.  When $q$ is a dominant strategy, 
$I(q,p(t))$ can never increase when $p(t)$ evolves according to the replicator equation.  
We can think of $I(q,p(t))$ as the information that the population has left to learn.  
Thus, evolution is analogous to a learning process---an analogy that in the field of artificial intelligence is exploited by evolutionary algorithms.
\item  
In Section \ref{markov} we consider continuous-time, finite-state Markov processes.  Here we have probability distributions on a finite set $X$ evolving according to a linear equation called the master equation.  In this case $I(p(t),q(t))$ can never increase.   Thus, if $q$ is a steady state solution of the master equation, both $I(p(t),q)$ and $I(q,p(t))$ are nonincreasing.  We can always write $q$ as the Boltzmann distribution for
some energy function $E \maps X \to \R$, meaning that
\[            q_i =  \frac{\exp(-E_i / k T)}{\displaystyle{\sum_{j \in X} \exp(-E_j / k T)} } \]
where $T$ is temperature and $k$ is Boltzmann's constant.  In this case, $I(p(t),q)$ is proportional to a difference of free energies:
\[    I(p(t),q) = \frac{F(p) - F(q)}{kT}   .\]
Thus, the nonincreasing nature of $I(p(t),q)$ is a version of the Second Law of
Thermodynamics.  In a companion paper \cite{Pollard}, we examine how this result generalizes to non-equilibrium steady states of `open Markov processes', in which probability can flow in or out of the set $X$.
\item 
Finally, in Section \ref{reaction_networks} we consider chemical reactions and other 
processes described by reaction networks.   In this context we have nonnegative real populations $P_i$ of entities of various kinds $i \in X$, and these population distributions evolve according to a nonlinear equation called the rate equation.  We can generalize relative information from probability distributions to populations by setting
\[        I(P,Q) = \sum_{i \in X} P_i \ln\left(\frac{P_i}{Q_i}\right) - \left(P_i - Q_i\right) . \]
The extra terms cancel when $P$ and $Q$ are both probability distributions, but they ensure that $I(P,Q) \ge 0$ for arbitrary populations.  If $Q$ is a special sort of steady state solution of the rate equation, called a complex balanced equilibrium, $I(P(t),Q)$ can never increase when $P(t)$ evolves according to the rate equation.
\end{itemize}

\section{The Replicator Equation}
\label{sec.replicator}

The replicator equation is a simplified model of how populations change with time.  Suppose we have $n$ different types of self-replicating entity.  We will call these entities {\bf \boldmath{replicators}}.  We will call the types of replicators {\bf \boldmath{species}}, but they do not need to be species in the biological sense.  For example, the replicators could be genes, and the types could be alleles.  Or the replicators could be restaurants, and the types could be restaurant chains.

Let $P_i(t),$ or just $P_i$ for short, be the population of the $i$th species at time $t.$  Then a very general form of the {\bf \boldmath{Lotka--Volterra equations}} says that
\beq \displaystyle{ \frac{d P_i}{d t} = f_i(P_1, \dots, P_n) \, P_i }. \label{lotka-volterra}
\eeq
Thus the population $P_i$ changes at a rate proportional to $P_i,$ but the `constant of proportionality' need not be constant: it can be any smooth function $f_i$ of the populations of all the species.  We call $f_i$ the {\bf \boldmath{fitness function}} of the $i$th species.   We can create a vector whose components are all the populations:
\[ P = (P_1, \dots , P_n). \]
This lets us write the Lotka--Volterra equations more concisely as
\[ \displaystyle{ \dot P_i = f_i(P) P_i} , \]
where the dot stands for a time derivative.

Instead of considering the population $P_i$ of the $i$th species, one often considers the probability $p_i$ that a randomly chosen replicator will belong to the $i$th species.   More precisely, this is the fraction of replicators belonging to that species:
\[ \displaystyle{  p_i = \frac{P_i}{\sum_j P_j} }. \]
As a mnemonic, remember that the \textbf{P}opulation $P_i$ is being normalized to give a \textbf{p}robability $p_i.$ 

How do these probabilities $p_i$ change with time?  The quotient rule gives
\[ \displaystyle{ \dot{p}_i = \frac{\dot{P}_i \left(\sum_j P_j\right) - P_i \left(\sum_j \dot{P}_j \right)}{\left(  \sum_j P_j \right)^2 } }\]
so the replicator equation gives
\[ \displaystyle{ \dot{p}_i = \frac{ f_i(P) P_i \; \left(\sum_j P_j\right) - P_i \left(\sum_j f_j(P) P_j \right)} {\left(  \sum_j P_j \right)^2 } .}\]
Using the definition of $p_i,$ this simplifies to:
\[ \displaystyle{ \dot{p}_i =  f_i(P) p_i  - \left( \sum_j f_j(P) p_j \right) p_i } .\]

The expression in parentheses here has a nice meaning: it is the {\bf \boldmath{mean fitness}}.  In other words, it is the average, or expected, fitness of a replicator chosen at random from the whole population.  Let us write it thus:
\[ \displaystyle{ \langle f(P) \rangle = \sum_j f_j(P) p_j  }. \]
This gives the {\bf \boldmath{replicator equation}} in its classic form:
\beq \displaystyle{ \dot{p}_i = \Big( f_i(P) - \langle f(P) \rangle \Big) \, p_i } \label{replicator} \eeq
where the dot stands for a time derivative. Thus, for the fraction of replicators of the $i$th species to increase, their fitness must exceed the mean fitness.   

So far, all this is classic material from population dynamics.  At this point, Marc Harper considers what information theory has to say \cite{Harper1,Harper2}.  For example, consider the relative information $I(q,p)$ where $q$ is some fixed probability distribution.  How does this change with time?  First, recall that
\[  I(q,p) = \displaystyle{ \sum_i  q_i  \ln \left(\frac{q_i}{ p_i }\right) }, \]
and we are assuming only $p_i$ depends on time, not $q_i$, so
\[  \displaystyle{ \frac{d}{dt} I(q,p(t))} =\displaystyle{ - \sum_i \frac{\dot{p}_i}{p_i} \, q_i } .\]
By the replicator equation we obtain
\[ \displaystyle{ \frac{d}{dt} I(q,p(t)) } = \displaystyle{ - \sum_i \Big( f_i(P) - \langle f(P) \rangle  \Big) \, q_i } . \]

This is nice, but we can massage this expression to get something more enlightening.  Remember, the numbers $q_i$ sum to one.  So:
\[ \begin{array}{ccl}  \displaystyle{ \frac{d}{dt} I(q,p(t)) } &=&  \displaystyle{  \langle f(P) \rangle - \sum_i f_i(P) q_i  } \\ \; \\ &=& \displaystyle{  \sum_i f_i(P) (p_i - q_i) . }
\end{array} \]
This result looks even better if we treat the numbers $f_i(P)$ as the components of a vector $f(P),$ and similarly for the numbers $p_i$ and $q_i.$  Then we can use the dot product of vectors to write
\beq \displaystyle{ \frac{d}{dt} I(q,p(t)) = f(P(t)) \cdot (p(t) - q) }  \label{dotI}, \eeq
whenever $p$ evolves according to the replicator equation while $q$ is fixed. It follows that the relative information $I(q,p(t))$ will be nonincreasing if and only if
\[ f(P(t)) \cdot (p(t) - q) \le 0 .\]
This nice result can be found in Marc Harper's 2009 paper relating the replicator equation to Bayesian inference \cite[Theorem 1]{Harper2}.  He traces its origins to much earlier work by Akin \cite{Akin1,Akin2}, and also Hofbauer, Schuster and Sigmund \cite{HofbauerSchusterSigmund}, who worked with a certain function of $I(q,p(t))$
rather than this function itself.
 
Next we turn to the question: how can we interpret the above inequality, and when does it hold?   

\section{Evolutionary Game Theory}
\label{game}

To go further, evolutionary game theorists sometimes assume the fitness functions
are linear in the probabilities $p_j$.  Then
\[ \displaystyle{f_i(P)   = \sum_{j=1}^n A_{ij} p_j } \]
for some matrix $A$, called the {\bf \boldmath{fitness matrix}}.   

In this situation the mathematics is connected to the usual von Neumann--Morgenstern theory of two-player games.  In this approach to game theory, each player has the same finite set $X$ of {\bf \boldmath{pure strategies}}.  The {\bf \boldmath{payoff matrix}} $A_{ij}$ specifies the first player's winnings if the first player chooses the pure strategy $i$ and the second player chooses the pure strategy $j$.  A probability distribution on the set of pure strategies is called a {\bf \boldmath{mixed  strategy}}.     The first player's expected winnings will be $p \cdot A q$ if they use the mixed strategy $p$ and the second player uses the mixed strategy $q$.  

To apply this analogy to game theory, we assume that we have a well-mixed population of replicators.   Each one randomly roams around, `plays games' with each other replicator it meets, and reproduces at a rate proportional to its expected winnings.   A pure strategy is just what we have been calling a species.   A mixed strategy is a probability distribution of species.  The payoff matrix is the fitness matrix.

In this context, the vector of fitness functions is given by
\[    f(P) =  A p .\]
Then, if $p(t)$ evolves according to the replicator equation while $q$ is fixed, the time derivative of relative information, given in Equation (\ref{dotI}), becomes
\beq  \displaystyle{ \frac{d}{dt} I(q,p(t)) = (p(t) - q) \cdot A p(t) } .  \label{replicator_information_gain}  \eeq
Thus, we define $q$ to be a {\bf \boldmath{dominant mixed strategy}} if 
\beq   q \cdot A p \ge p \cdot A p \label{dominance} \eeq 
for all mixed strategies $p$.  If $q$ is dominant, we have
\beq \displaystyle{ \frac{d}{dt} I(q,p(t)) \le 0 }  \label{dominance_information_gain} \eeq
whenever $p(t)$ obeys the replicator equation.  Conversely, if the information of 
$p(t)$ relative to $q$ is nonincreasing whenever $p(t)$ obeys the replicator equation, 
the mixed strategy $q$ must be dominant.

The question is then: what is the meaning of dominance?   First of all,  if $q$ is dominant then it is a {\bf \boldmath{steady state}} solution of the replicator equation, meaning one that does not depend on time.  To see this, let $r(t)$ be the solution of the replicator equation with $r(0) = q$.  Then $I(q,r(t))$ is nonincreasing because $q$ is dominant.  Furthermore $I(q,r(t)) = 0$ at $t = 0$, since for any probability distribution we have $I(q,q) = 0$.  Thus we have $I(q,r(t)) \le 0$ for all $t \ge 0$.  However, relative information is always non-negative, so we must have $I(q,r(t)) = 0$ for all $t \ge 0$.  This forces $r(t) = q$, since the relative information of two probability distributions can only vanish if they are equal.

Thus, a dominant mixed strategy is a special kind of steady state solution of the replicator equation.  But what is special about it?  We can understand this if we think in game-theoretic terms.   The inner product $q \cdot A p$ will be my expected winnings if I use the mixed strategy $q$ and you use the mixed strategy $p$.   Similarly, $p \cdot A p$
will be my expected winnings if we both use the mixed strategy $p$.   So, the dominance of the mixed strategy $q$ says that my expected winnings can never increase if I switch from $q$ to whatever mixed strategy you are using.

It helps to set these ideas into the context of evolutionary game theory \cite{Sandholm}.  In 1975, John Maynard Smith, the founder of evolutionary game theory, defined a mixed strategy $q$ to be an `evolutionarily stable state' if when we add a small population of `invaders' distributed according to any other probability distribution $p,$ the original population is more fit than the invaders \cite{MaynardSmith1}.  He later wrote \cite{MaynardSmith}: ``A population is said to be in an evolutionarily stable state if its genetic composition is restored by selection after a disturbance, provided the disturbance is not too large.'' 

More precisely, Maynard Smith defined $q$ to be an {\bf \boldmath{evolutionarily stable state}} if 
\[ q \cdot A ((1-\epsilon)q + \epsilon p) >  p \cdot A ((1-\epsilon)q + \epsilon p) \]
for all mixed strategies $p \ne q$ and all sufficiently small $\epsilon > 0 .$   Here 
\[ (1-\epsilon)q + \epsilon p \]
is the population we get by replacing an $\epsilon$-sized portion of our original population by invaders.  

Taking this inequality and separating out the terms of order $\epsilon$, one can easily check that $q$ is an evolutionarily stable state if and only if two conditions hold for 
all probability distributions $p \ne q$:
\beq q \cdot A q \ge p \cdot A q \label{ess1}   \eeq
and
\beq q \cdot A q = p \cdot A q \; \Rightarrow \; q \cdot A p > p \cdot A p \label{ess2}. \eeq
The first condition says that $q$ is a {\bf \boldmath{symmetric Nash equilibrium}}.  In other words, the invaders cannot on average do better playing against the original population than members of the original population are.    The second says that if the invaders are just as good at playing against the original population, they must be worse at playing against each other!  The combination of these conditions means the invaders won't take over.  

Note, however, that the dominance condition (\ref{dominance}), which guarantees nonincreasing relative information, is different from either Equation (\ref{ess1}) or 
(\ref{ess2}).    Indeed, after Maynard Smith came up with his definition of `evolutionarily stable state', Bernard Thomas \cite{Thomas} came up with a different definition.   For him, $q$ is an {\bf \boldmath{evolutionarily stable strategy}} if Maynard Smith's condition (\ref{ess1}) holds along with
\beq q \cdot A p \ge p \cdot A p . \label{Thomas2} \eeq
This condition is stronger than Equation (\ref{ess2}), so he renamed Maynard Smith's evolutionarily stable states {\bf \boldmath{weakly  evolutionarily stable strategies}}.  

More importantly for us, Equation (\ref{Thomas2}) is precisely the same as the condition we are calling `dominance', which implies that the relative information $I(q,p(t))$ can never increase as $p(t)$ evolves according to the replicator equation.  We can interpret $I(q,p(t))$ as the amount of information `left to learn' as the population approaches the dominant strategy.

This idea of evolution as a learning process is exploited by genetic algorithms in artificial intelligence \cite{Mitchell}.  Conversely, some neuroscientists have argued that individual organisms act to minimize `surprise'---that is, relative information: the information of perceptions relative to predictions \cite{FristonAo}.  As we shall see in the next section, relative information also has the physical interpretation of free energy.   Thus, this hypothesis is known as the `free energy principle'.  Another hypothesis, that neurons develop in a manner governed by natural selection, is known as `neural Darwinism' \cite{Edelman}.  The connection between relative information decrease and evolutionary game theory shows that these two hypotheses are connected.

\section{Markov Processes}
\label{markov}

One limitation of replicator equations is that in these models, when the population of some species is initially zero, it must remain so for all times.  Thus, they cannot model mutation, horizontal gene transfer, or other sources of novelty.  

The simplest model of mutation is a discrete-time Markov chain, where there is a fixed probability per time for a genome to change from one genotype to another each time it is copied \cite{Nielsen}.  The information-theoretic aspects of Markov models in genetics have been discussed by Sober and Steel \cite{SoberSteel}.  To stay within our overall framework, here we instead consider continuous-time Markov chains, which we shall simply call Markov processes.  These are a very general framework that can be used to describe any system with finite set $X$ of states where the probability that the system is in its $i$th state obeys a differential equation of this form:
\[          \frac{dp_i}{dt} = \sum_{j \in X} H_{ij} p_j(t)  \]
with the matrix $H$ chosen so that total probability is conserved.  

In what follows we shall explain a very general result saying that for any Markov process, relative information is nonincreasing \cite{Gorban,LieseVajda, Moran}.  It is a form of the Second Law of Thermodynamics.  Some call this result the `$H$-theorem', but this name goes back to Boltzmann, and strictly this name should be reserved for arguments like Boltzmann's which seek to derive the Second Law from time-symmetric dynamics together with time-asymmetric initial conditions \cite{Price,Zeh}.  The above equation is not time-symmetric, and the relative information decrease holds for all initial conditions.

We can describe a Markov process starting with a directed graph whose nodes correspond to states of some system, and whose edges correspond to transitions between these states.  The transitions are labelled by `rate constants', like this:
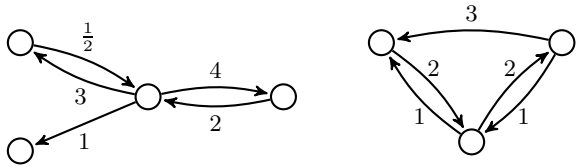
\[
\begin{tikzpicture}[->,>=stealth',shorten >=1pt,thick,scale=0.6]
  \node[main node] (1) at (0,2.2) {};
  \node[main node](2) at (0,-.2) {};
  \node[main node](3) at (2.83,1)  {};
  \node[main node](4) at (5.83,1) {};
  \node[main node](4a) at (8,2.2) {};
  \node[main node](5) at (10,0) {};
  \node[main node](6) at (12,2.2) {};
  \path[every node/.style={font=\sffamily\small}, shorten >=1pt]
    (3) edge [bend left=12] node[above] {$4$} (4)
    (4) edge [bend left=12] node[below] {$2$} (3)
    (3) edge [bend left=0] node[below] {$1$} (2)
    (1) edge [bend left=12] node[above] {$\frac{1}{2}$}(3) 
    (3) edge [bend left=12] node[below] {$3$} (1)
    (4a) edge [bend left=12] node[above] {$2$} (5)
    (5) edge [bend left=12] node[below] {$1$} (4a) 
    (5) edge [bend left=12] node[above] {$2$} (6)
    (6) edge [bend right=12] node[above] {$3$}(4a) 
    (6) edge [bend left=12] node[below] {$1$} (5);
\end{tikzpicture}
\]
The rate constant of a transition from $i \in X$ to $j \in X$ represents the probability per time that an item hops from the $i$th state to the $j$th state.

More precisely, we say a {\bf \boldmath{Markov process}} $M$ consists of:
\begin{itemize}
\item a finite set $X$ of {\bf \boldmath{states}}, 
\item a finite set $T$ of {\bf \boldmath{transitions}}, 
\item maps $s,t \maps T \to X$ assigning to each transition its {\bf \boldmath{source}} and {\bf \boldmath{target}},  
\item a map $r \maps T \to (0,\infty)$ assigning a {\bf \boldmath{rate constant}} $r(\tau)$ to each transition $\tau \in T$. 
\end{itemize}
 If $\tau \in T$ has source $i$ and target $j$, we write $\tau \maps i \to j$. 

From a Markov process we can construct a square matrix, or more precisely a 
function $H \maps X \times X \to \R$, called its {\bf \boldmath{Hamiltonian}}.  
If $i \ne j$ we define
\[ H_{i j} = \sum_{\tau \maps j \to i} r_\tau  \]
to be the sum of the rate constants of all transitions from $j$ to $i$.   We choose the diagonal entries in a more subtle way:
\[  H_{i i} = -\sum_{\substack{\tau \maps i \to j \\ j \ne i}} r(\tau). \]

Given a Markov process, the {\bf \boldmath{master equation}} for a time-dependent probability distribution on $X$ is:
\beq    \frac{d}{dt}p(t) = Hp(t)  \label{master}, \eeq
where $H$ is the Hamiltonian.  Thus, given a probability distribution $p$ on $X$, for $i \ne j$ we interpret $H_{ij}p_j $ as the rate at which population flows from state $j$ to state $i$, while the quantity $H_{ii}p_i$ is the outflow of population from state $i$.  The diagonal entries $H_{ii}$ are chosen in a way that ensures total population is conserved.   

More precisely, $H$ is {\bf \boldmath{infinitesimal stochastic}}, meaning that its off-diagonal entries are non-negative and the entries in each column sum to zero:
\[ H_{ij} \geq 0 \; \textrm{ if } i \neq j \quad \textrm{and} \quad \sum_i H_{ij} = 0.\]  
This guarantees that if $p(t)$ obeys the master equation and if it is initially a probability distribution, it remains a probability distribution for all times $t \ge 0$.  

Markov processes are an extremely general formalism for dealing with randomly evolving systems, and they are presented in many different ways in the literature.  For example, besides the master equation one often sees the {\bf \boldmath{Kolmogorov forward equation}}: 
\[   \frac{d}{dt} G(t,s) = HG(t,s)    \]
where $G(t,s)$ is a square matrix depending on two times $s,t \in \R$ with $s \le t$.  The idea here is that the matrix element $G_{ij}(t,s)$ is the probability that if the system is in the $j$th state at time $s$, it will be in the $i$th state at some later time $t$.  We thus demand that $G(t,s)$ is the identity matrix when $s = t$, and we  can show that
\[        G(t,s) = \exp((t-s)H) \]
whenever $s \le t$.   From this it is easy to see that $G(t,s)$ also obeys the
{\bf \boldmath{Kolmogorov backward equation}}:
\[    \frac{d}{ds} G(t,s) = -G(t,s)H .\]
We should warn the reader that conventions differ and many, perhaps even most, authors multiply these matrices in the reverse order.

The master equation and Kolmogorov forward equation are related as follows.  If $p(t)$ obeys the master equation and $G(t,s)$ solves the Kolmogorov forward 
equation, then 
\[        p(t) = G(t,s) p(s), \]
whenever $s \le t$.   Thus, knowledge of $G(t,s)$ immediately tells us all solutions of the master equation.

Most of our discussion so far, and the results to follow, can be generalized to the case where $X$ is an arbitrary measure space, for example $\R^n$.  The Kolmogorov forward equation is often studied in this more general context, sometimes in the guise of the `Fokker--Planck equation'.  This formulation is often used to study Brownian motion and other random walk processes in the continuum.  A careful treatment of this generalization involves more analysis: sums become integrals, and one needs to worry about convergence and passing derivatives through integrals \cite{EthierKurtz,Norris,RogersWilliams1,RogersWilliams2}.   To keep things simple and focus on basic concepts, we only treat the case where $X$ is a finite set.  

As one evolves any two probability distributions $p$ and $q$ according to a Markov process, their relative information is nonincreasing:
\[   \frac{d}{dt}  I(p(t),q(t))  \le 0.  \]
This is a very nice result, because it applies regardless of the Markov process.  It even applies to a master equation where the Hamiltonian depends on time, as long as it is always infinitesimal stochastic.

To prove this result, we start by computing the derivative:
\[ \begin{array}{ccl} \displaystyle{ \frac{d}{dt}  I(p(t),q(t)) } &=& 
\displaystyle{ \frac{d}{dt} \sum_{i \in X} p_i \ln(\frac{p_i}{q_i}) } \\  \\
&=& \displaystyle{ \sum_{i} \dot{p}_i \ln(\frac{p_i}{q_i}) + p_i\left(\frac{\dot{p}_i}{p_i} - \frac{\dot{q}_i}{q_i} \right) } \\  \\
&=&  \displaystyle{ \sum_{i,j} H_{ij} p_j  \ln(\frac{p_i}{q_i}) + p_i\left(\frac{H_{ij}p_j}{p_i} - \frac{H_{ij}q_j}{q_i} \right) }
\end{array} \]
where in the second line we used the master equation.   We can rewrite this as
\[    \displaystyle{ \frac{d}{dt}  I(p(t),q(t)) =  \sum_{i,j} H_{ij} p_j  \left( \ln(\frac{p_i}{q_i}) + 1 - \frac{p_i q_j}{p_j q_i} \right). } \]
Note that the last two terms cancel when $i = j$.  Thus, if we break the sum into an $i \ne j$ part and an $i = j$ part, we obtain
\[    \displaystyle{ \frac{d}{dt}  I(p(t),q(t)) =  \sum_{i \ne j} H_{ij} p_j  \left( \ln(\frac{p_i}{q_i}) + 1 - \frac{p_i q_j}{p_j q_i} \right) + \sum_j H_{jj} p_j \ln(\frac{p_j}{q_j}). } \]
Next we use the infinitesimal stochastic property of $H$ to write $H_{jj}$ as the
sum of $-H_{ij}$ over $i$ not equal to $j$:
\[ \begin{array}{ccl}   \displaystyle{ \frac{d}{dt}  I(p(t),q(t))}  &=& \displaystyle{ \sum_{i \ne j} H_{ij} p_j  \left( \ln(\frac{p_i}{q_i}) + 1 - \frac{p_i q_j}{p_j q_i} \right) - \sum_{i \ne j} H_{ij} p_j \ln(\frac{p_j}{q_j}) }  \\ \\
&=& \displaystyle{ \sum_{i \ne j} H_{ij} p_j  \left( \ln(\frac{p_iq_j}{p_j q_i}) + 1 - \frac{p_i q_j}{p_j q_i} \right). }
\end{array} \]
Since $H_{ij} \ge 0$ when $i \ne j$ and $\ln(s) + 1 - s \le 0$ for all $s > 0$, we
conclude that
\beq   \frac{d}{dt} I(p(t),q(t)) \le 0   \label{second_law}  \eeq
as desired.  To be precise, this derivation only applies when $q_i$ is nonzero for all $i \in X$.  If this is true at any time, it will be true for all later times.  If some probability $q_i$ vanishes, the relative entropy $I(p,q)$ can be infinite.  As we evolve $p$ and $q$ in time according to the master equation, the relative entropy can drop from infinity to a finite value, but never increase.

One of the nice features of working with a finite state space $X$ is that in this case
every Markov process admits one or more {\bf \boldmath{steady states}}: probability distributions $q$ that obey 
\[             H q = 0 \]
and thus give solutions of the master equation that are constant in time \cite{BB}.   If we fix any one of these, we can conclude
\beq      \frac{d}{dt} I(q,p(t)) \le 0 \label{second_law_1}  \eeq
for any solution of the master equation.   This is the same inequality we have already seen for the replicator equation when $q$ is a dominant mixed strategy, namely Equation (\ref{dominance_information_gain}).  But for a Markov process, we also have
\beq      \frac{d}{dt} I(p(t),q) \le 0 ,    \label{second_law_2} \eeq
and this, it turns out, has a nice meaning in terms of statistical mechanics.

In statistical mechanics we want to assign an {\bf \boldmath{energy}} $E_i$ to each state such that the steady state probabilities $q_i$ are given by the so-called {\bf \boldmath{Boltzmann distribution}}:
\beq                     q_i = \frac{e^{-\beta E_i}}{Z(\beta)} .  \label{Boltzmann}  \eeq
Here $\beta$ is a parameter which in physics is defined in terms of the temperature $T$ by $\beta = 1/kT$, where $k$ is Boltzmann's constant.  The quantity $Z(\beta)$ is a normalizing constant called the {\bf \boldmath{partition function}}, defined by
\beq                   Z(\beta) = \sum_{i \in X} e^{-\beta E_i}   \label{partition} \eeq
to ensure that the probabilities $q_i$ sum to one.  

However, whenever we have a probability distribution $q$ on a finite set $X$, we can turn this process on its head.  We start by arbitrarily choosing $\beta > 0$.  Then we define energy differences by
\beq             E_i - E_j = -\beta^{-1} \ln(\frac{q_i}{q_j})  . \label{energies} \eeq
This determines the energies up to an additive constant.  If we make a choice for these energies, we can define the partition function by Equation (\ref{partition}), and 
Boltzmann's law, Equation (\ref{Boltzmann}), will follow.  

We can thus apply ideas from statistical mechanics to any Markov process, for example the process of genetic drift.  The concepts of `energy' and `temperature' play only a metaphorical role here; they are not the ordinary physical energy and temperature.  However, the metaphor is a useful one.

So, let us fix a Markov process on a set $X$ together with a steady state probability distribution $q$.   Let us choose a value of $\beta$, choose energies obeying Equation (\ref{energies}), and define the partition function $Z(\beta)$ by Equation (\ref{partition}).
To help the reader's intuition we define a {\bf \boldmath{temperature}} $T = 1/\beta$, setting Boltzmann's constant to 1.   Then, for any probability distribution $p$ on $X$ we can define the {\bf \boldmath{expected energy}}:
\[         \langle E \rangle_p = \sum_{i \in X} p_i E_i  \]
and the {\bf \boldmath{entropy}}:
\[               S(p) = - \sum_{i \in X} p_i \ln(p_i) . \]
From these, we can construct the all-important {\bf \boldmath{free energy}} 
\[              F(p) = \langle E \rangle_p - T S(p) . \]
In applications to physics and chemistry this is, roughly speaking, the amount of 
`useful' energy, meaning energy not in the form of random `heat', which gives the
term $T S(p)$.

We can prove that this free energy can never increase with time if we evolve $p$ in time
according to the master equation.  This is a version of the Second Law of Thermodynamics. To prove this, note that
\[ \begin{array}{ccl}    F(p) &=& \displaystyle{ \langle E \rangle_p - T S(p)}  \\  \\
&=&  \displaystyle{ \sum_{i \in X} p_i E_i + T \,p_i \ln(p_i),  }
\end{array} \]
but by Boltzmann's law, Equation (\ref{Boltzmann}), we have
\[    E_i = - T \left(\ln(q_i) + \ln(Z) \right)  \]
so we obtain
\[     F(p) = - T \sum_{i \in X} \left( p_i \ln(q_i) - p_i \ln(p_i) + p_i \ln(Z)\right),  \]
or, using the definition of relative information and the fact that the $p_i$ sum to one:
\[      F(p) =  T(I(p,q) - \ln(Z))  .\]
In the special case where $p = q$ the relative information vanishes and we obtain
\[      F(q) = - T \ln(Z) .\]
Substituting this into the previous equation, we reach an important result:
\beq      \frac{F(p) - F(q)}{T} = I(p,q)     \label{free_energy_as_relative_entropy} .\eeq
Relative information is proportional to a difference in free energies!  Since relative entropy is nonnegative, we immediately see that any probability distribution $p$ has at least as much free energy as the steady state $q$:
\[            F(p) \ge F(q)   .\]
Moreover, if we evolve $p(t)$ according to the master equation, the decrease of 
relative entropy given by Equation (\ref{second_law_2}) implies that
\[    \frac{d}{dt}  F(p(t)) \le 0 .\]
These two facts suggest, but do not imply, that $p(t) \to q$ as $t \to \infty$.  This is in fact true when there is a unique steady state, but not necessarily otherwise.   One can determine the number of linearly independent steady states from the topology of the graph associated to the Markov process \cite[Section 22.2]{BB}. 

\section{Reaction Networks}
\label{reaction_networks}

Reaction networks are commonly used in chemistry, where they are called `chemical reaction networks'. An example is the Krebs cycle, important in the metabolism of aerobic organisms.  A simpler example is the Michaelis--Menten model of an enzyme $E$ binding to a substrate $S$ to form an intermediate $I$, which in turn can break apart into a product $P$ and the original enzyme:

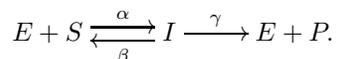
\[
\xymatrix{
E + S \ar@<0.6ex>[r]^<<<<<{\alpha} 
& I   \ar@<0.6ex>[l]^<<<<<{\beta}  \ar[r]^<<<<<{\gamma} 
& E + P.
}
\]
Mathematically, this is a directed graph.  The nodes of this graph, namely $E + S, I$, and $E + P$,  are called `complexes'.  Complexes are finite sums of `species', which in this example are $E, S, I,$ and $P$.  The edges of this graph are called `reactions'.   Each reaction has a name, which may also serve as the `rate constant' of that reaction. In real-world chemistry, every reaction has a reverse reaction going the other way, but if the rate constant for the reverse reaction is low enough, we may simplify our model by omitting it.  This is why the Michaelis--Menten model has no reaction going from $E + P$ back to $I$.   

From a reaction network we can extract a differential equation called its `rate equation', which describes how the population of each species changes with time.  We treat these populations as functions of time, taking values in $[0,\infty)$.  If we use $P_E$ as the name for the population of the species $E$, and so on, the rate equation for the above reaction network is:
\[ 
\begin{array} {ccl}
\vspace{6pt} 
\displaystyle{\frac{d}{dt}} P_E &=& - \alpha P_E P_S + \beta P_I + \gamma P_I \\ \vspace{6pt} 
\displaystyle{\frac{d}{dt}} P_S &=& - \alpha P_E P_S + \beta P_I  \\  \vspace{6pt} 
\displaystyle{\frac{d}{dt}} P_I &=& \alpha P_E P_S - \gamma P_I  \\  \vspace{6pt} 
\displaystyle{\frac{d}{dt}} P_P &=& \gamma P_I
\end{array}
\]
We will give the general rules for extracting the rate equation from a reaction network, but the reader may enjoy guessing them from this example.   It is worth noting that chemists usually deal with `concentrations' rather than populations: a concentration is a population per unit volume.   This changes the meaning and the values of the rate constants, but the mathematical formalism is the same.

More precisely, a {\bf \boldmath{reaction network}} consists of:
\begin{itemize}
\item a finite set $S$ of {\bf \boldmath{species}},
\item a finite set $X$ of {\bf \boldmath{complexes}} with $X \subseteq \N^S$,
\item a finite set $T$ of {\bf \boldmath{reactions}} or {\bf \boldmath{transitions}},
\item maps $s,t \maps T \to X$ assigning to each reaction its {\bf \boldmath{source}} and {\bf \boldmath{target}}, 
\item a map $r \maps T \to (0,\infty)$ assigning to each reaction a {\bf \boldmath{rate constant}}.
\end{itemize}
The reader will note that this looks very much like our description of a Markov process
in Section \ref{markov}.  As before, we have a graph with edges labelled by rate constants.   However, now instead of the nodes of our graphs being abstract `states', they are complexes: 
finite linear combinations of species with natural number coefficients, which we can write as elements of $\mathbb{N}^S$.   

For convenience we shall often write $k$ for the number of species
present in a reaction network, and identify the set $S$ with the set
$\{1, \dots, k\}$.  This lets us write any complex as a $k$-tuple of
natural numbers.  In particular, we write the source and target of any
reaction $\tau$ as
\[    s(\tau) = (s_1(\tau), \dots, s_k(\tau)) \in \N^k , \]
\[    t(\tau) = (t_1(\tau), \dots,  t_k(\tau)) \in \N^k . \]

The rate equations involve the {\bf \boldmath{population}} $P_i \in [0,\infty)$ of each species $i$.
We can summarize these in a population vector 
\[      P = (P_1, \dots , P_k)  .\]
The rate equations say how this vector change with time.  It 
says that each reaction $\tau$ contributes to the time derivative of $P$ 
via the product of:
\begin{itemize}
\item the vector $t(\tau) - s(\tau)$ whose $i$th component is the change
in the number of items of the $i$th species due to the reaction $\tau$;
\item the concentration of each input species $i$ of $\tau$ raised to
the power given by the number of times it appears as an input, namely
$s_i(\tau)$;
\item the rate constant $r(\tau)$ of $\tau$.
\end{itemize}
The {\bf \boldmath{rate equations}} are
\beq
\frac{d}{dt} P(t) =
\sum_{\tau \in T} r(\tau) (t(\tau) - s(\tau)) P(t)^{s(\tau)},
\label{rate_equation}
\eeq
where $P \maps \R \to [0,\infty)^k$ and we have
used multi-index notation to define
\[  P^{s(\tau)} = P_1^{s_1(\tau)} \cdots P_k^{s_k(\tau)}. \]
Alternatively, in components, we can write the rate equation as
\[
\dot{P}_i = 
\sum_{\tau \in T} r(\tau) (t_i(\tau) - s_i(\tau)) P^{s(\tau)} .
\]
The reader can check that this rule gives the rate equations for the Michaelis--Menten model.  

Reaction networks include Markov processes as a special case.    A reaction network where every complex is just a single species---that is, a vector in $\N^S$ with one component being $1$ and all the rest $0$---can be viewed as a Markov process.    For a reaction network that corresponds to a Markov process in this way, the rate equation is linear, and it matches the master equation for the corresponding Markov process.  The goal of this section is to generalize results on relative information from Markov processes to other reaction networks.  However, the nonlinearity of the rate equation introduces some subtleties.

The applications of reaction networks are not limited to chemistry.  Here is an example that arose in work on  HIV, the human immunodeficiency virus \cite{Kor}:
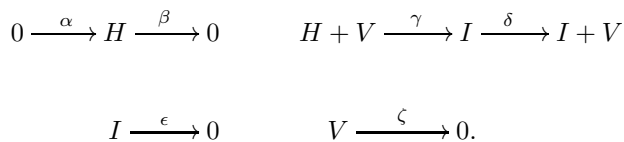
\[
\xymatrix{
0 \ar[r]^\alpha & H \ar[r]^{\beta} & 0  & 
H + V \ar[r]^<<<<<\gamma & I \ar[r]^>>>>>>\delta & I + V 
\\  & I  \ar[r]^\epsilon &0  &  V  \ar[r]^\zeta &0 .
}
\]
Here we have three species:
\begin{itemize}
\item $H$: healthy white blood cells,
\item $I$:  infected white blood cells, 
\item $V$: virions (that is, individual virus particles).
\end{itemize}
The complex $0$ above is short for $0 H + 0 I + 0 V$: that is, `nothing'.  
We also have six reactions:
\begin{itemize}
\item $\alpha$: the birth of one healthy cell, which has no input and
one $H$ as output.
\item $\beta$: the death of a healthy cell, which has one $H$ as input
and no output.
\item $\gamma$: the infection of a healthy cell, which has one $H$ and
one $V$ as input, and one $I$ as output.
\item $\delta$: the reproduction of the virus in an
infected cell, which has one $I$ as input, and one $I$ and one $V$ as
output.
\item $\epsilon$: the death of an infected cell, which has one $I$ as
input and no output.
\item $\zeta$: the death of a virion, which has one $V$ as input and
no output.
\end{itemize}
For this reaction network, if we use the Greek letter names of the reactions as names for their rate constants, we get these rate equations:
\[ \begin{array}{ccl}
\vspace{6pt} 
\displaystyle{ \frac{d}{dt} P_H } &=& \alpha - \beta P_H - \gamma P_H P_V
\\ \vspace{6pt} 
\displaystyle{ \frac{d}{dt}} P_I &=&   \gamma P_H P_V - \epsilon P_I
\\ \vspace{6pt} 
\displaystyle{ \frac{d}{dt}} P_V &=& - \gamma P_H P_V + \delta P_I  - \zeta P_V.
\end{array}
\]

The equations above are not of the Lotka--Volterra type shown in Equation (\ref{lotka-volterra}), because the time derivative of $P_H$ contains a term with no factor of $P_H$, and similarly for $P_I$ and $P_V$.   Thus, even when the population of one of these three species is initially zero, it can become nonzero.  However, many examples of Lotka--Volterra equations do arise from reaction networks.  For example, we could take two species:
\begin{itemize}
\item $R$: rabbits,
\item $W$: wolves,
\end{itemize}
and form this reaction network:
\[
\xymatrix{
R \ar[r]^\alpha & 2R  & R + W  \ar[r]^<<<<<\beta & 2 W  & W \ar[r]^\gamma &0 .
}
\]
Taken literally, this seems like a ludicrous model: rabbits reproduce asexually, 
a wolf can eat a rabbit and instantly give birth to another wolf, and wolves can also die.
However, the resulting rate equations are a fairly respectable special case of the famous Lotka--Volterra predator-prey model:
\[ \begin{array}{ccl}
\vspace{6pt} 
\displaystyle{ \frac{d}{dt} P_R } &=& \alpha P_R - \beta P_R P_W 
\\  \vspace{6pt} 
\displaystyle{ \frac{d}{dt}} P_W &=&  \beta P_R P_W - \gamma P_W.
\end{array}
\]
It is probably best to think of this as saying no more than this: general results about reaction networks will also apply to Lotka--Volterra equations that can arise from this framework.

In our discussion of the replicator equation, we converted populations to probability
distributions by normalizing them, and defined relative information only for the resulting probability distributions.  We can, however, define relative information for populations, and this is important for work on reaction networks.  Given two populations $P,Q \maps X \to [0,\infty)$, we define
\beq \displaystyle{ I(P,Q) = \sum_{i \in X} P_i \ln(\frac{P_i}{Q_i}) - (P_i - Q_i) } .
\label{relative_information_for_populations} \eeq
When $P$ and $Q$ are probability distributions on $X$ this reduces to the relative information defined before in Equation (\ref{relative_information}).   As before, one can prove that 
\[       I(P,Q) \ge 0 .\]
To see this, note that a differentiable function $f \maps \R \to \R$ is convex precisely when its graph lies above any of its tangent lines:
\[      f(y) \ge f(x) + f'(x) (y - x) .\]
This is true for the exponential function, so
\[      e^y \ge e^x + e^x(y - x)  \]
and thus for any $p,q > 0$ we have 
\[        q \ge p + p(\ln(q) - \ln(p))  \]
or
\[        p \ln(\frac{p}{q}) - (p - q) \ge 0 . \]
Thus, each term of the sum in Equation (\ref{relative_information_for_populations}) is greater than or equal to zero, so $I(P,Q) \ge 0$.  Furthermore since we have equalities above only when $x = y$, or in other words $p = q$, we also obtain
\[      I(P,Q) = 0 \iff P = Q  .\]
So, relative information has the properties of a divergence, but for arbitrary populations $P,Q \maps X \to [0,\infty).$

A function very similar to $I(P,Q)$ was used by Friedrich Horn and Roy Jackson in their important early paper on reaction networks \cite{HornJackson}.  They showed that this function is nonincreasing when $P$ evolves according to the rate equation and $Q$ is a steady state of a special sort, called a `complex balanced equilibrium'.  Later Martin Feinberg, another of the pioneers of reaction network theory, gave a shorter proof of this fact \cite{Feinberg}. Our goal here is  to explain this result and present Feinberg's proof.

We say that a population $Q$ is {\bf \boldmath{complex balanced}} if
for each complex $\kappa \in K$ we have
\beq  \displaystyle{ \sum_{\tau: s(\tau) = \kappa} r(\tau) Q^{s(\tau)} = 
\sum_{\tau: t(\tau) = \kappa} r(\tau) Q^{s(\tau)}   } . 
\label{complex_balance}
\eeq
This says that each \emph{complex} is being produced at the same rate at which it is being destroyed.   This is stronger than saying $Q$ is a steady state solution of the rate equation.  On the other hand, it is weaker than the `detailed balance' condition saying that each reaction occurs at the same rate as the reverse reaction.  The founders of chemical reaction theory discovered that many results about detailed balanced equilibria can just as easily be shown in the complex balanced case.   The calculation below is an example.

We have
\[ \frac{d}{dt} I(P(t), Q) = \sum_{i \in X} \dot{P}_i \ln(\frac{P_i}{Q_i}),   \]
so, using the rate Equation (\ref{rate_equation}), we obtain:
\[ \begin{array}{ccl}
 \displaystyle{ \frac{d}{dt} I(P(t), Q) }&=&\displaystyle{ \sum_{i \in X} \sum_{\tau \in T} 
r(\tau) \big(t_i(\tau) - s_i(\tau)\big) \ln(\frac{P_i}{Q_i}) \; P^{s(\tau)} } \\  \\
&=& \displaystyle{ \sum_{i \in X} \sum_{\tau \in T} 
r(\tau) \left[ \ln\left(\left(\frac{P_i}{Q_i}\right)^{t_i(\tau)}\right) - 
\ln \left(\left(\frac{P_i}{Q_i}\right)^{s_i(\tau)}\right)\right]  P^{s(\tau)} } .\\ 
\end{array}
\]
We can convert each sum over $i$ of the logarithms into a logarithm of a product, and if we define a vector
\[       \frac{P}{Q} = ( \frac{P_1}{Q_1}, \dots, \frac{P_k}{Q_k} ),\]
we can use multi-index notation to write these products very concisely, 
obtaining
\[ \begin{array}{ccl}    \displaystyle{ \frac{d}{dt} I(P(t), Q) } 
&=&  \displaystyle{
\sum_{\tau \in T} r(\tau) \left[ \ln\left(\left(\frac{P}{Q}\right)^{t(\tau)}\right) - \ln \left(\left(\frac{P}{Q}\right)^{s(\tau)}\right)\right]  P^{s(\tau)} } \\  \\
&=&  \displaystyle{
\sum_{\tau \in T} r(\tau) \left[ 
\ln\left(\left(\frac{P}{Q}\right)^{t(\tau)}\right) - 
\ln \left(\left(\frac{P}{Q}\right)^{s(\tau)}\right)     \right] 
\left(\frac{P}{Q}\right)^{s(\tau)}  Q^{s(\tau)}. } 
\end{array}
\]
Then, using the fact that $(\ln x - \ln y) y \le x - y$, we obtain
\[   \displaystyle{ \frac{d}{dt} I(P(t), Q) } \le 
 \displaystyle{ \sum_{\tau \in T} 
r(\tau) \left[ \left(\frac{P}{Q}\right)^{t(\tau)} \! - 
\left(\frac{P}{Q}\right)^{s(\tau)} \right]  Q^{s(\tau)}  } . \]
Next, we write the sum over reactions as a sum over complexes $\kappa$ and then a sum over reactions having $\kappa$ is their target (for the first term) or target (for the second):
\[    \displaystyle{ \frac{d}{dt} I(P(t), Q) } \le 
\sum_{\kappa \in X} \left[  \sum_{\tau: t(\tau) = \kappa} 
r(\tau)  \left(\frac{P}{Q}\right)^{t(\tau)} Q^{s(\tau)}   - \sum_{\tau: s(\tau) = \kappa}  r(\tau)  \left(\frac{P}{Q}\right)^{s(\tau)} Q^{s(\tau)} 
\right].  \]
We can pull out the factors involving $\frac{P}{Q}$:
\[    \displaystyle{ \frac{d}{dt} I(P(t), Q) } \le 
\sum_{\kappa \in X} \left(\frac{P}{Q}\right)^{\kappa} \left[  \sum_{\tau: t(\tau) = \kappa} 
r(\tau)   Q^{s(\tau)}   - \sum_{\tau: s(\tau) = \kappa}  r(\tau)   Q^{s(\tau)}
\right],  \]
but now the right side is zero by the complex balanced condition, Equation (\ref{complex_balance}).  Thus, we have
\beq    \frac{d}{dt} I(P(t), Q) \le 0    \label{h-theorem_for_reaction_networks}, \eeq
whenever $P(t)$ evolves according to the rate equation and $Q$ is a complex balanced equilibrium.  

As noted above, a reaction network where every complex consists of
a single species gives a linear rate equation.  In this special case we can strengthen the above result: we have
\beq    \frac{d}{dt} I(P(t) , Q(t)) \le 0    \label{h-theorem_for_linear_reaction_networks} \eeq
whenever $P(t)$ and $Q(t)$ evolve according to the rate equation.  The reason is
in this case, the rate equation is also the master equation for a Markov process.  Thus, we can reuse the argument leading up to inequality (\ref{second_law}) for
Markov processes, since nothing in this argument used the fact that the probability distributions were normalized. 

\section{Conclusions}
\label{conclusions}

We have seen theorems guaranteeing that relative information cannot increase in three different situations: evolutionary games described by the replicator equation, Markov processes, and reaction networks.   In all cases, the decrease of relative entropy is closely connected to the approach to equilibrium as $t \to \infty$.  For the replicator equation, this equilibrium is a dominant mixed strategy.  For a Markov process, whenever there is a unique steady state, all probability distributions approach this steady state as $t \to \infty$.   For reaction networks, the appropriate notion of equilibrium is a complex balanced equilibrium, generalizing the more familiar concept of detailed balanced equilibrium.   

It is natural to inquire about the mathematical relation between these results. Inequality (\ref{h-theorem_for_linear_reaction_networks}) for Markov processes resembles inequality (\ref{h-theorem_for_reaction_networks}) for reaction networks.  However, neither result subsumes the other.  The master equation for a Markov process is a special case of the rate equation for a reaction network.  However, the result for reaction networks says only that 
\[      \displaystyle{  \frac{d}{dt}  I(P(t),Q) \le 0 } \]
when $Q$ is a complex balanced equilibrium and $P(t)$ obeys the rate
equation, while the result for Markov processes says that
\[      \displaystyle{  \frac{d}{dt}  I(P(t),Q(t)) \le 0  }  \]
whenever $P(t)$ and $Q(t)$ obey the master equation.
Furthermore, neither of these inequalities subsume or are subsumed by the result for the replicator equation, inequality (\ref{dominance_information_gain}).   Indeed, this result applies only to the probability distributions obtained by normalizing population distributions,  not populations.  Furthermore it is `turned around', in that sense that
$q$ appears first:
\[         \displaystyle{  \frac{d}{dt}  I(q,p(t)) \le 0  } \]
whenever $p(t)$ obeys the replicator equation and $q$ is a dominant strategy.  We know of no results showing that $I(p(t),q)$ is nonincreasing when $p(t)$ obeys the replicator equations, nor results showing that $I(P(t),Q)$ or $I(Q,P(t))$ is nonincreasing when $P(t)$ obeys the Lotka--Volterra equation. 

In short, while relative entropy is nonincreasing in the approach to equilibrium in 
all three situations considered here, the details differ in significant ways.  A challenging open problem is thus to find some `super-theorem' that has all three of these results as special cases.   The work of Gorban \cite{Gorban} is especially interesting in this regard, since he tackles the challenge of finding new nonincreasing functions for reaction networks.

\subsection*{Acknowledgments}

We thank the referees for significant improvements to this paper.
We thank Manoj Gopalkrishnan \cite{Gopalkrishnan} for explaining his proof that $I(P(t),Q)$ is nonincreasing when $P(t)$ evolves according to the rate equation of a reaction network and $Q$ is a complex balanced equilibrium.  We thank David Anderson \cite{Anderson} for explaining Martin Feinberg's proof of this result.  We also thank NIMBioS, the National Institute for Mathematical and Biological Synthesis, for holding a workshop on Information and Entropy in Biological Systems at which we presented some of this work.

\end{document}